\documentclass[conference]{IEEEtran}
\IEEEoverridecommandlockouts
\usepackage{cite}
\usepackage{amsmath,amssymb,amsfonts}
\usepackage{algorithmic}
\usepackage{amsmath,graphicx}
\usepackage[caption=false]{subfig}

\usepackage{textcomp}
\usepackage{xcolor}
\def\BibTeX{{\rm B\kern-.05em{\sc i\kern-.025em b}\kern-.08em
    T\kern-.1667em\lower.7ex\hbox{E}\kern-.125emX}}
\usepackage{booktabs} 

\usepackage{etoolbox}
\makeatletter
\patchcmd{\@makecaption}
  {\scshape}
  {}
  {}
  {}
\makeatletter
\patchcmd{\@makecaption}
  {\\}
  {.\ }
  {}
  {}
\makeatother
\def\tablename{Table}

\usepackage{booktabs,caption}
\usepackage[flushleft]{threeparttable}

\begin{document}

\title{Toward end-to-end interpretable convolutional neural networks for waveform signals}

\author{
    \IEEEauthorblockN{Linh Vu}
    \IEEEauthorblockA{\textit
        \textit{Monash University}\\
        linh.vu@monash.edu}
    
    \and   

    \IEEEauthorblockN{Thu Tran}
    \IEEEauthorblockA{\textit
        \textit{Singapore Management University}\\
        ndttran.2019@phdcs.smu.edu.sg}
    \and
    
    \IEEEauthorblockN{Lim Wern Han}
    \IEEEauthorblockA{\textit
        \textit{Monash University}\\
        lim.wern.han@monash.edu}
    \and
    
    \IEEEauthorblockN{Rapha\"{e}l C.-W. Phan}
    \IEEEauthorblockA{\textit
        \textit{Monash University}\\
        raphael.phan@monash.edu}
}

\maketitle

\begin{abstract}
  This paper introduces a novel convolutional neural networks (CNN) framework tailored for end-to-end audio deep learning models, presenting advancements in efficiency and explainability. By benchmarking experiments on three standard speech emotion recognition datasets with five-fold cross-validation, our framework outperforms Mel spectrogram features by up to seven percent. It can potentially replace the Mel-Frequency Cepstral Coefficients (MFCC) while remaining lightweight. Furthermore, we demonstrate the efficiency and interpretability of the front-end layer using the PhysioNet Heart Sound Database, illustrating its ability to handle and capture intricate long waveform patterns. Our contributions offer a portable solution for building efficient and interpretable models for raw waveform data.

\end{abstract}


\section{Introduction}
\label{sec:introduction}
The use of deep learning and representation learning has significantly impacted the signal processing field, similar to the success observed in the fields of Natural Language Processing (NLP) and Computer Vision (CV). Deep neural networks can automatically learn hierarchical representations from raw data. However, simply using deep learning as a one-tool-fit-all solution often leads to poor interpretability, i.e., monitoring how the model works is difficult. 

\begin{figure}
    \includegraphics[width=\columnwidth]{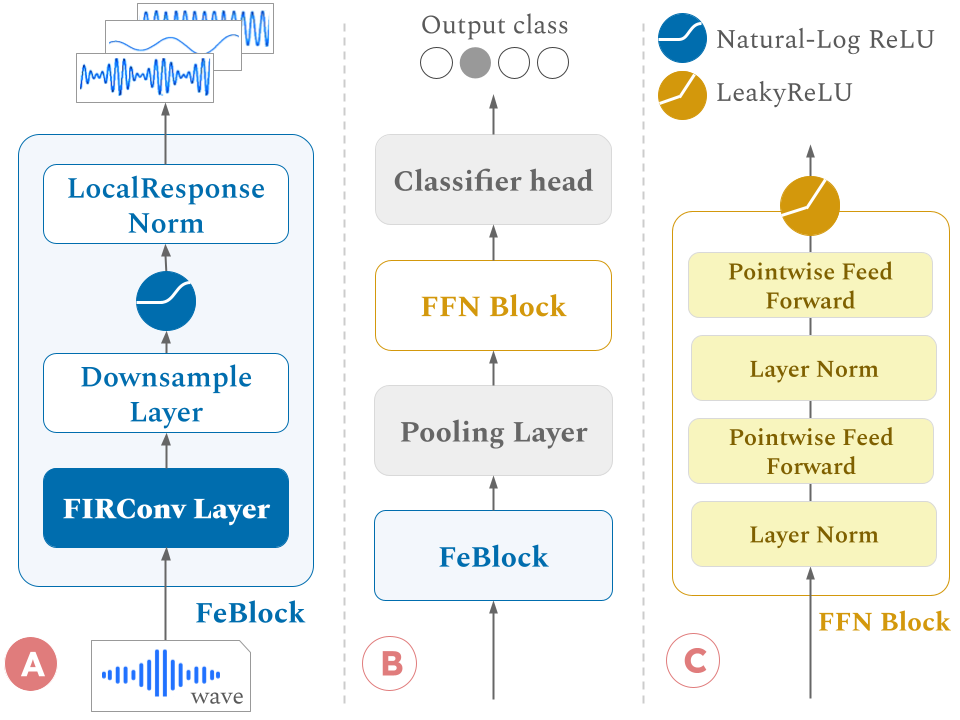}
    \caption[IConNet]{The proposed IConNet architecture for end-to-end audio classification: \textit{A}- the front-end block containing the FIRconv layer; \textit{B}- the proposed general architecture for end-to-end audio classification; \textit{C}- the classifier used in the experiments.}
    \label{fig:iconnet}
\end{figure}

Recent developments have led to the creation of efficient and easy-to-understand models for processing raw waveform signals. Cheuk et al. \cite{cheuk2020nnaudio} have introduced nnAudio, a framework that uses 1D convolutional neural networks to extract spectrograms on-the-fly. Leiber et al. \cite{leiber2022differentiable_stft_window} have introduced a differentiable modification of the Short-Time Fourier Transform (STFT) that makes it possible to optimize the window length parameter through gradient descent. These works primarily revolve around rethinking the fusion of spectrograms and neural networks, placing emphasis on employing an adapted spectrogram as the initial layer for input signals. Seki et al. \cite{seki2017deep} showed that Gaussian filters with three trainable parameters of gain, bandwidth, and centre frequency can outperform Mel-filterbank in both clean and noise-corrupted environments. Inspired by standard filtering in digital signal processing, Ravanelli and Bengio \cite{ravanelli2019interpretable} proposed a novel CNN called SincNet that uses rectangular band-pass filters as kernels for the first convolution layer with two learnable parameters: high and low cut-off frequencies. Right from the first layer, SincNet can already learn meaningful filters and thus converge faster than the conventional CNN. In a comparison between different parametric modulated kernel-based filters including SincNet (rectangular filters), Sinc2Net (triangular filters), GammaNet (gammatone filters) and GaussNet (Gaussian filters); it is found that SincNet is the best for the speech recognition task \cite{Loweimi2019}.

According to Ravanelli and Bengio \cite{ravanelli2019interpretable}, the frequency gain of the SincNet filters is only defined in subsequent layers of the neural network, which are conventional CNN layers. To further contribute to this research direction, we propose a new interpretable CNN architecture called IConNet that utilizes a finite impulse response (FIR)-based kernel with learnable window functions. FIR filters play a crucial role in signal processing as they enable the extraction of critical information; and different filter designs can enhance or mitigate unwanted effects. With adaptive window functions, the filters can dynamically adjust to varying signal profiles depending on the data and the problem. The primary benefit of this approach is the transparency in the way the model learns -- which frequency bands it focuses on and which will be cut off. We illustrate the effectiveness of this approach in two health-related problems: speech emotion recognition and abnormal heart sound detection.

In the next section \ref{sec:method}, we will describe the proposed method. Section \ref{sec:ser} is about the Speech Emotion Recognition (SER) experiment, followed by section \ref{sec:heart} about abnormal heart sound detection.

\section{The IConNet architecture} 
\label{sec:method}

The proposed Interpretable Convolutional Neural Network (IConNet) architecture in this research is designed to leverage insights from audio signal processing; aiming to improve end-to-end deep neural networks' ability to extract features and patterns from raw waveform signals. The foundation of our method draws inspiration from the standard signal processing process, in which the input audio signal undergoes a windowing operation to segment it into smaller frames and simultaneously mitigate spectral leakage to improve frequency resolution. A key novelty lies in using the Generalized Cosine Window function as parametrization for the convolution kernels to enable the neural networks to choose the most suitable shape for each frequency band. 

The convolution layer of the proposed model has restricted-shaped kernels which is a band-pass filter defined by non-learnable low cut-off frequency $f_0$ and frequency bandwidth $f_\delta$. The filter shape and frequency gain are determined by the window function $W$ with $p$ learnable parameters $\phi_p$. Let $H = \{h_k: \quad k=1, ..., K\}$ denotes the kernel (filter) of width $K$ in the time domain. $V_n$ denotes the output of the convolution layer for each $n$ input time-domain value. $H(k, f_0, f_\delta, \phi)$ is parametrically modulated by $sinc$ as a non-learnable function and $W$ with learnable parameters.

\begin{equation}
    V_n = X \cdot H = \sum_{k=1}^{K}X_k\cdot h_{n-k}
\end{equation}

\begin{equation}
    H = T * W = T(k, f_0, f_\delta) * W(k, \phi), \qquad 0 \leq k < K
\end{equation}

\begin{multline}
    T(k, f_0, f_\delta) = \displaystyle{2(f_0+f_\delta){sinc}(2{\pi}k(f_0+f_\delta))} \\ 
\displaystyle{- 2{f_0}{sinc}(2{\pi}k{f_0}), \qquad sinc(a) = \sin(a)/a}
\end{multline}

\begin{equation}    
    W(k, \phi) = \biggl\{w_k \Big| w_k = \sum_{i=0}^{p}(-1)^i\phi_i\cos{\frac{2{\pi}ik}{K-1}}\biggr\}  
\end{equation}

Thus, for each kernel, there are only $p$ parameters $\phi_p$ and two band parameters $f_0$ and $f_\delta$ to train via gradient descent. $f_0$ and $f_\delta$ can also be non-learnable, which is how we can incorporate prior domain knowledge into the deep neural network design to solve specific problems and control what information is fed into the model to prevent unwanted effects from the first layers.

After separating signals into different channels, we use a downsampling layer to reduce the data in the time domain to a lower sample rate. This compression step decreases the data's dimensions, making it easier to process in subsequent steps while retaining the essential features needed for precise analysis and classification. It's worth noting that the convolution filters from the previous step also serve as anti-aliasing filters like traditional signal downsampling. In the second-to-last step, we use an NLRelu activation function recommended in \cite{liu2019naturallogrelu} to mimic the amplitude-to-decibel conversion since human hearing functions on a logarithmic scale.  Finally, we apply the Local Respond Normalization function \cite{krizhevsky2012imagenet} to the output before forwarding it to the next layer. This entire process can be repeated by stacking these blocks on top of each other to achieve further pattern decorrelation from previous decomposed channels while maintaining a more compressed representation. The outputs of the front-end blocks can be combined together as long as they are resampled to the same sampling rate to preserve the temporal characteristic. Depending on the task, these front-end blocks can be incorporated into any deep neural network architecture. 

Figure \ref{fig:iconnet} illustrates the proposed architecture, with the part \textbf{\textit{A}} on the left describing the front-end block. The middle part \textbf{\textit{B}} of the figure \ref{fig:iconnet} is the high-level deep neural network architecture that incorporates the front-end block to process raw waveform signal data. Part \textbf{\textit{C}} depicts a simple classifier block consisting of a pooling layer, a two-layer feed-forward neural network with layer norm and a LeakyRelu activation function. The following experiments are designed to demonstrate the feature extraction ability of the front-end blocks and evaluate the effectiveness of this new architecture. 

We have selected two classification problems and designed a classifier consisting of a pooling layer, a two-layer feed-forward neural network with layer norm and a LeakyRelu activation function. In both experiments, we use k-fold cross-validation settings and report three different metrics: unweighted accuracy (UA), unweighted F1 (UF1) and weighted-F1 (F1). For training the neural networks, we employed RAdam optimizer \cite{liu2019radam} with OneCycleLR learning rate scheduler \cite{smith2017onecyclelr}, Cross Entropy loss and trained each model to up to 60 epochs on each fold.

\section{Speech Emotion Recognition} 
\label{sec:ser}

\subsection{Background}
Emotions play a pivotal role in revealing essential cues about a speaker's intentions, attitudes, and mental state, particularly in the context of AI health. This significance is underscored by a recent study from Gheorghe et al. \cite{gheorghe2023dnn_depression} where a system leveraging deep neural networks successfully identifies depression from speech samples with an unweighted accuracy of 91.25\% using Mel-frequency cepstral coefficients (MFCC). MFCC is the most popular feature based on the short-time Fourier Transform spectrogram and the Mel filterbank, which was designed based on human perception of our hearing system. As humans perceive sound on a logarithmic scale, Mel filterbank uses narrower bandwidths at the lower frequencies to capture more information. Its filters have a triangular shape and are used in many tasks. However, according to \cite{sarangi2020optimization, seki2017deep}, in the data-driven filterbanks, each filterbank's centre frequencies and shapes are different depending on the tasks and the presence of background noise.

Based on these insights, we've developed three IConNet variants detailed in Table \ref{table:ser_models}. These variants assess the influence of adjustable bands versus adaptive windows on CNN model classification outcomes. The front-end kernels in all IConNet models are initialized with the Mel-filterbank, with additional filters allocated to the lower frequency range, prioritizing crucial speech signals.

\subsection{Experiment setup}

\subsubsection{Datasets}
Our first experiment incorporates the proposed method, evaluating its performance on three standard Speech Emotion Recognition (SER) datasets with similar settings described in \cite{vu2022muser} \footnote{It is worth mentioning that in \cite{vu2022muser}, Vu et al. benchmarked the performance of handcrafted features after 500 epochs on the 20\% test set after 10-fold cross validation on 80\% of the train set.}. Firstly, the \textbf{RAVDESS} dataset \cite{livingstone2018ryerson} features high-quality audio speech with eight emotional expressions from 24 North-American actors. Secondly, the \textbf{CREMA-D} dataset \cite{cao2014crema} comprises recordings from 91 speakers with diverse races and ethnicities, representing real-world audio recordings where the recordings are often noise-corrupted. Lastly, \textbf{IEMOCAP} \cite{busso2008iemocap} is a popular SER dataset with 12-hour conversational speech audio from 10 speakers. We use four classes of data, namely \emph{happy, sad, angry, neutral}, which are shared between the aforementioned three datasets for a fair comparison.

\subsubsection{Classifiers and evaluation}
For the classifier, we use a dense neural network consisting of two layers with 512 nodes in each layer followed by layer norm, as described in the diagram \ref{fig:iconnet}. To evaluate the performance of the proposed models, we adopted 5-fold cross-validation with stratified train-test splitting to ensure the proportion of each class in both sets. We trained each model for up to 60 epochs with early stopping, then evaluated the model on the test set and reported the average metrics across folds. We benchmarked IConNet with different settings against Mel and MFCC features extracted using the \textit{TorchAudio} library \cite{yang2021torchaudio} with different resolutions as described in the table \ref{table:ser_models}. 

\begin{table}[!ht]
    \centering
	\caption[SER models]{Models used in the SER experiment}
    \label{table:ser_models}
    \begin{tabular}{cl}
        \toprule
        \textbf{Model} & \textbf{Description} \\
        \midrule
        \textbf{Mel}-\textit{K} & Mel spectrogram with \textit{K} Mel frequency bins \\
        \textbf{MFCC}-\textit{K} & MFCC from Mel-\textit{K} \\
        \midrule
        \textbf{IConNet-B}-\textit{K} & IConNet with \textit{K} learnable-band filters\\
        \textbf{IConNet-W}-\textit{K} & IConNet with \textit{K} learnable-window filters\\
        \textbf{IConNet-BW}-\textit{K} & IConNet with \textit{K} learnable-band-window filters\\
        \bottomrule
    \end{tabular}
\end{table}

\subsection{Experiment results}
\begin{table*}[!ht]
    \begin{threeparttable}
        \centering
        \caption[Results for each dataset]{Results of hand-crafted features and IConNet on three datasets in percentage with early stopping (\%)}
        \label{table:ser_result_each_dataset}
        \begin{tabular}[width=\linewidth]{l|ccc|ccc|ccc|ccc}
            \toprule
            \textbf{Dataset} & &  \textbf{RAVDESS} &  &  & \textbf{CREMA-D} &  &  & \textbf{IEMOCAP} &  & &   \textbf{Average} &  \\ 
            \midrule
            \textbf{Model} & \textbf{UA} & \textbf{UF1} & \textbf{F1} & \textbf{UA} & \textbf{UF1} & \textbf{F1} & 
            \textbf{UA} & \textbf{UF1} & \textbf{F1} &  
            \textbf{UA} & \textbf{UF1} & \textbf{F1} \\ 
            \midrule
                
            \textbf{Mel-256} &
            51.59 &	51.73 &	53.22 &	
            59.07 &	59.07 &	58.83 &	
            50.85 &	50.85 &	51.13 &	
            54 ±5	& 54 ±5	& 54 ±4 \\ 
    
            \textbf{Mel-456} &
            52.00 &	49.09 &	51.67 &	
            59.98 &	59.98 &	59.60 &	
            52.85 &	52.85 &	53.03 &	
            55 ±4	& 54 ±6	& 55 ±4 \\ 
                
            \midrule
                
            \textbf{MFCC-256} &
            49.29 &	46.01 &	47.58 &	
            56.47 &	56.27 &	56.70 &	
            \textbf{\underline{56.68}} &	\textbf{\underline{56.41}} &	\textbf{\underline{56.25}} &	
            54 ±4	& 53 ±6	& 54 ±5 \\ 
    
            \textbf{MFCC-456} &
            45.21 &	42.27 &	46.13 &	
            56.70 &	56.57 &	56.86 &	
            56.34 &	56.07 &	56.14 &	
            53 ±7	& 52 ±8	& 53 ±6 \\
                
            \midrule
                
            \textbf{IConNet-B-256} & 
            63.73 &	63.40 &	65.01 &	
            61.67 &	61.31 &	61.53 &	
            53.72 &	53.37 &	53.37 &	
            60 ±5	& 59 ±5	& 60 ±6 \\ 
    
            \textbf{IConNet-B-456} &
            62.00 &	62.46 &	64.65 &	
            61.40 &	61.67 &	61.94 &	
            53.17 &	53.12 &	53.51 &	
            59 ±5	& 59 ±5	& 60 ±6 \\ 
    
            \midrule
                
            \textbf{IConNet-W-256} &
            \textbf{65.83} &	\textbf{65.02} &	\textbf{66.65} &	
            \textbf{62.30} &	\textbf{62.30} &	\textbf{62.06} &	
            56.44 &	55.94 &	56.17 &	
            \textbf{62 ±5}	& \textbf{61 ±5}	& \textbf{62 ±5} \\ 
    
            \textbf{IConNet-W-456} &
            \textbf{\underline{66.83}} &	\textbf{\underline{66.15}} &	\textbf{\underline{67.37}} &	
            \textbf{\underline{62.34}} &	\textbf{\underline{62.34}} &	\textbf{\underline{62.56}} &	
            \textbf{56.67} &	\textbf{56.32} &	\textbf{\underline{56.56}} &	
            \textbf{\underline{62 ±5}}	& \textbf{\underline{62 ±5}}	& \textbf{\underline{62 ±5}} \\ 
    
            \midrule
                
            \textbf{IConNet-BW-256} &
            64.84 &	64.83 &	66.35 &	
            61.08 &	61.08 &	60.75 &	
            52.78 &	52.77 &	53.37 &	
            60 ±6	& 60 ±6	& 60 ±7 \\ 
    
            \textbf{IConNet-BW-456} &
            62.36 &	62.07 &	63.73 &	
            61.28 &	61.28 &	61.00 &	
            53.26 &	53.76 &	54.21 &	
            59 ±5	& 59 ±5	& 60 ±5 \\ 
                
            \bottomrule
        \end{tabular} 
        \begin{tablenotes}
          \small
          \item Notation: \textbf{\underline{bold-underline}}: best results, \textbf{bold}: second-best results
          
        \end{tablenotes}
    \end{threeparttable}
    \end{table*}

\begin{figure}[t]
\includegraphics[width=1\linewidth]{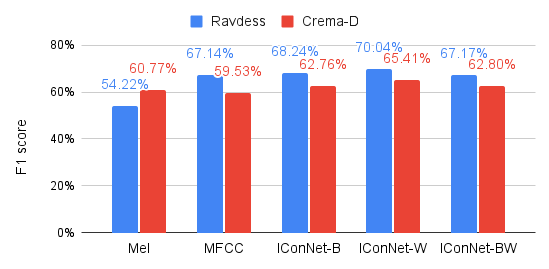}
\caption{Result on RAVDESS and CREMA-D datasets after 60 epochs}
\label{fig:ser_60epochs}
\end{figure}

\begin{figure}[t]
\includegraphics[width=1\linewidth]{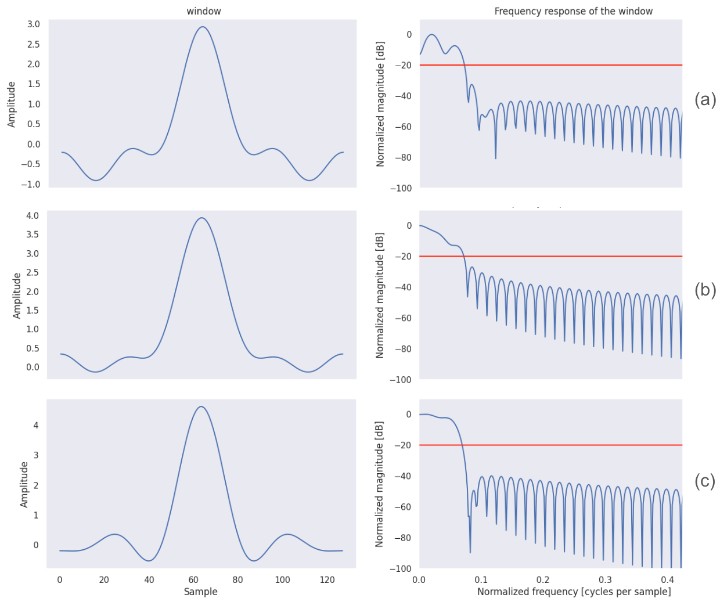}
\caption{Comparison of Window Shape and Frequency Response of Filters from Different Bands. The chart displays the frequency response of low-range (a), mid-range (b), and high-range (c) frequency bands. The red line at -20dB represents the threshold at which noise is perceived as not noticeable.}
\label{fig:ser_windows}
\end{figure}

Table \ref{table:ser_result_each_dataset} provides detailed experiment result on the three datasets mentioned above when using early stopping. On the RAVDESS dataset, the IConNet model with 456 adaptive-window FIR kernels achieved an unweighted accuracy of 66.83\%, which is 4.83\% higher than the adjustable-band-FIR model with the same number of kernels. When the number of kernels was reduced to 256, the UA of the IConNet-W dropped 1\%, while the IConNet-B increased 1.73\%. For the IConNet-W models and Mel-spectrogram models, higher resolution helped the models make better predictions. The size of Mel-256/IConNet-256 and Mel-456/IConNet-456 models is 1.6 MB and 2 MB, respectively. Despite having the same size, Mel models performed poorly compared to all IConNet models, even after being trained for more epochs.

The bar chart \ref{fig:ser_60epochs} compares the F1 scores of Mel-456, MFCC-456, and IConNet-456 models on the RAVDESS and CREMA-D datasets. After 60 epochs, the performance of MFCC on the RAVDESS dataset increased from 46.13\% to 67.14\%, which is a more than 21\% improvement. However, the gain for MFCC-456 model on CREMA-D is less than 1\%. The IConNet-W model still gave the highest F1 score on both the RAVDESS and CREMA-D datasets at 70.04\% and 65.41\%, respectively.

The table in \ref{table:ser_result_each_dataset} clearly indicates that IConNet-W-456 and MFCC-256 attained the highest unweighted accuracy on IEMOCAP with only a 0.01\% difference between them. On average, the IConNet-W models outperformed other IConNet variants by roughly 2\%. Overall, IConNet models gave better results than Mel-spectrogram and MFCC models, especially on the CREMA-D dataset. 

On the interpretable ability of the IConNet, Figure \ref{fig:ser_windows} demonstrates the alterations in the shape of windows that are used to extract important information from different frequency ranges. The window \textit{(a)} is learned during the model training process for the low-frequency range and exhibits a complex shape, determining the frequency gain at each point. This complex shape enhances the robustness of the model because the low-frequency range contains crucial speech signals that are susceptible to noise. On the other hand, the window \textit{(c)} is tailored for the higher frequency range, serving as a narrow-band filter that effectively mitigates spectral leakage with its ideal sidelobes.

In summary, the experiment results have proven the appropriate use of the proposed IConNet, especially the IConNet with adaptive-window FIR kernel for SER. Moreover, the IConNet end-to-end models are portable, with the model size being 30\% smaller than the hand-crafted feature set models proposed by Vu et al. in \cite{vu2022muser}.

\section{Abnormal heart sound detection} 
\label{sec:heart}
\subsection{Background}
Cardiovascular diseases are a leading global health concern, demanding improved diagnostic precision. A \textit{heart murmur} is an unusual sound heard during the heartbeat cycle, indicating underlying cardiac conditions. Various ML approaches have been developed to identify heart murmurs from stethoscope recordings. As the recording duration is quite long (30 seconds on average), there are often many preprocessing steps, such as heartbeat segmentation to trim the waveform to a smaller size (3 to 5 seconds) and bandpass filtering as the most critical signals lay in the range from 10 Hz to 400 Hz. After that, MFCC and its derivatives are extracted to apply CNN or LSTM models or a combination of both \cite{li2021lightweight, li2022heart}.

\subsection{Proposed model}
This experiment mainly examines the potential role of the IConNet in refining abnormal heart sound detection. Our proposed model is an end-to-end lightweight neural network architecture that eliminates the need for heart sound segmentation, low-pass filtering and MFCC-based feature extraction. We designed an IConNet-W model with two front-end blocks with 128 and 32 kernels respectively, a max-pooling layer and a 2-layer FFN classifier with 256 nodes on each layer. The total number of parameters is below 200K.

\subsection{Experiment setup}
We employ the widely-used \textbf{PhysioNet/CinC Challenge} dataset \cite{liu2016physionet} for heart sound classification evaluation. This dataset comprises 2575 \textit{normal} and 665 \textit{abnormal} samples. To validate the effectiveness of the IConNet in identifying relevant features, we resample the waveform from 2000 Hz to 16000 Hz and conduct 4-fold cross-validation, reporting UA, UF1, and F1 metrics. Deng et al.\cite{deng2020heart} serve as our baseline, utilizing a preprocessing pipeline with a bandpass filter, MFCC features, and a CRNN model consisting of three 2D CNN layers and two LSTM layers. We also include the MFCC performance for comparison. Preprocessing for other models involves waveform trimming and downsampling, excluding the IConNet model.

\subsection{Experiment results}
Based on the results presented in Table \ref{tab:result_heartsound}, it is clear that the baseline model \cite{deng2020heart} performed better than the MFCC + FFN model, thanks to its preprocessing steps that included band-pass filtering and the use of MFCC deltas. The baseline model achieved 90.06\% F1 score. However, our proposed architecture surpassed both models with an F1 score of 92.05\%, which is 2\% higher than the baseline model. While this result still not yet outperforms the state-of-the-art Resnet result reported by Li et al. in \cite{li2022heart}, it has successfully demonstrated the effectiveness of our proposed method in classifying heart sound data. Furthermore, the visualization of the front-end filters confirms that it allocates band-pass filters that actively change the window shapes to extract essential information in the range of 643 ±134 Hz. The windows have learned to transform into band-stop filter shapes for the high-frequency range above 2000 Hz, which only contain meaningless artifacts from the resampling step. Understanding the features utilized by the backbone neural network model, which influences its decisions, is vital for ensuring reliable outcomes, particularly in health applications.

\begin{table}[th]
  \caption{Abnormal heart sound detection result on the \textbf{PhysioNet} dataset}
  \label{tab:result_heartsound}
  \centering
  \begin{tabular}{ r@{}l c c c}
    \toprule
    \multicolumn{2}{c}{\textbf{Model}} & 
    \multicolumn{1}{c}{\textbf{UA}} & 
    \multicolumn{1}{c}{\textbf{UF1}} & 
    \multicolumn{1}{c}{\textbf{F1}} \\
    \midrule
    MFCC + FFN & & 82.98 & 83.97 & 88.68 \\
    MFCC deltas + CRNN \cite{deng2020heart} & & 85.67 & 80.21 & 90.60 \\
    IConNet & & 87.48 & 81.12 & 92.05 \\
    \bottomrule
  \end{tabular}
\end{table}

\begin{figure}[t]
\includegraphics[width=1\linewidth]{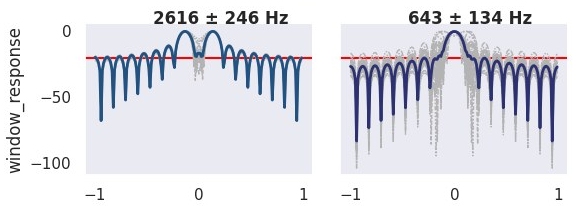}
\caption{Frequency response of filters from different bands}
\label{fig:heart_win}
\end{figure}

\section{Conclusions}
\label{sec:conclusions}

Our proposed framework introduces a novel method for constructing end-to-end audio deep-learning models, showcasing its efficacy in healthcare applications such as speech emotion recognition and abnormal heart sound detection. Our findings reveal that the proposed CNN framework surpasses traditional methods utilizing the Mel spectrogram, and potentially MFCC (further experimentation required for confirmation), for both tasks. Moreover, visualization of CNN kernels underscores the value of transparent CNN architectures in healthcare settings, shedding light on the features extracted from input signals in mission-critical tasks.

Insightful observations arise regarding the front-end layer within our proposed framework. Our results indicate that front-end layers featuring learnable windows demonstrate superior performance compared to those employing learnable bands, diverging from the predominant focus in existing literature. We suggest exploring methods to incorporate prior knowledge into the front-end layer to improve performance, especially with the learnable window model. While the proposed framework offers promising results, it requires slightly more parameters and, thus, more training resources.

In conclusion, this study emphasizes the advantages of employing interpretable CNN for analyzing raw waveform data, showcasing their potential benefits in healthcare settings to foster transparency and trust in medical AI applications.

\bibliographystyle{IEEEtran}

\bibliography{main}

\end{document}